\documentstyle[12pt]{article}

\begin{document}

\title{The Operator Algebra of the  Quantum Relativistic Oscillator}

\author{Ion I. Cot\u aescu\\ Gheorghe Dr\u ag\u anescu\\{\small \it The West 
University of Timi\c soara,}\\ {\small \it V. P\^ arvan Ave. 4, RO-1900 
Timi\c soara, Romania}}

\date{\today}

\maketitle

\begin{abstract}
The operator algebras of a new family of relativistic geometric models 
of the relativistic oscillator \cite{COTA} are studied. It is shown that, 
generally, the operator of number of quanta and the pair of the shift operators 
of each  model are the generators of  a non-unitary representation of the 
$so(1,2)$ algebra, except a special case when this algebra becomes the standard one of the  
non-relativistic harmonic oscillator. 

Pacs:  04.62.+v, 03.65.Ge
\end{abstract}

\section{Introduction}
\

In the general relativity, the geometric models play the role of  
kinematics, helping us to understand the characteristics of the  classical or 
quantum free motion on a given background. 
One of the simplest (1+1) geometric models is that of the  
(classical or quantum) relativistic harmonic oscillator (RHO). 
Based on  phenomenological \cite{P1} and group theoretical 
\cite{A1,A2} arguments, this has been  defined as a free system on the  
anti-de Sitter  static background. There exists a (3+1) anti-de Sitter static 
metric \cite{P1} which can be restricted to  the (1+1) metric 
given  by the line element 
\begin{equation}\label{(m1)}
ds^{2}=\frac{1}{1-\omega^{2} x^{2}}dt^{2}-\frac{1}{(1-\omega^{2}
x^{2})^{2}}dx^{2}.
\end{equation}
This metric reproduces the classical equation of motion of the non-relativistic 
harmonic oscillator (NRHO) of the frequency $\omega$. Moreover, the corresponding 
quantum model, represented by a free scalar massive quantum field, has an 
equidistant discrete spectrum with a ground state energy larger, but 
approaching $\omega/2$ in the non-relativistic limit (in natural units $\hbar 
=c=1$) \cite{M}. In a previous article \cite{COTA} we have generalized this 
model to the family of models depending on a real parameter $\lambda$ which 
has the metrics given by
\begin{equation}\label{(m)}
ds^{2}=g_{00}dt^{2}+g_{11}dx^{2}=
\frac{1+(1+\lambda) \omega^{2}x^{2}}{1+\lambda \omega^{2} x^{2}}dt^{2}-  
\frac{1+(1+\lambda) \omega^{2}x^{2}}{(1+\lambda \omega^{2} x^{2})^{2}}dx^{2}.
\end{equation} 
Here, the parametrization has been defined 
in order to obtain the exact anti-de Sitter metric (\ref{(m1)})  
for $\lambda = -1$. We have shown \cite{COTA}  that the quantum models 
with $\lambda>0$ have mixed energy spectra, with a finite discrete sequence and 
a continuous part, while for $\lambda\le 0$ these spectra are countable. 
However, despite of their  different relativistic behavior, 
all these models have the same non-relativistic limit, namely the NRHO. 
For this reason we shall use the name of relativistic 
oscillators (RO) for all the models with $\lambda \not= -1$, understanding 
that the RHO is only that of the anti-de Sitter metric.

In this article, we intend to study the operator algebra of the quantum RO with 
countable energy spectra. For these models $\lambda \le 0$ and, therefore, it 
is convenient to  denote 
\begin{equation}\label{(par2)}
\lambda=-\epsilon^{2}, \qquad \hat\omega =\epsilon \omega,
\end{equation} 
and to rewrite our previous results in this new notation.  
We take $\epsilon\ge 0$ so that the pure anti-de Sitter RHO will have 
$\epsilon=1$.

The energy spectra of our quantum models depend on only one quantum number 
\cite{COTA}. Therefore, we will have one operator of number of quanta and a 
pair of shift (i.e. raising or lowering) operators.  Our main objective 
is to identify the algebra which is linearly generated by  these operators. 
To this end, 
first we show that there exists a natural (holonomic) frame where the 
relativistic energy eigenfunctions coincide with those of the non-relativistic 
P\" oschl-Teller (PT) system \cite{PT1, PT2}. By using their known properties, 
we shall construct the main operators which act on the one-particle wave 
functions, giving a  special attention to the three mentioned ones. The result 
is that for all the RO with $\epsilon>0$, including the RHO, these   
are the generators of the $so(1,2)$ algebra which has as Casimir operator  
just the Klein-Gordon equation. In the limit $\epsilon\to 0$ the $so(1,2)$ 
algebra degenerates in the standard algebra of the NRHO.

We  start, in the second section, with a short review of the properties of 
the  Hilbert space of one-particle states  of the free quantum scalar massive 
field  in the coordinate representation. In the next section we  
briefly present our previous results concerning the RO, giving  the  energy 
levels and the energy eigenfunctions with their normalization factors. 
In the Sec. 4 we  show that, in the special frame where the metric is the 
conformal transformation of the Minkowski flat metric,  
the RO generates  a relativistic PT problem for which the shift operators of 
the energy basis are known \cite{PT2}. Based on these results we  analyze, 
in the next section, the operator algebra of the models with $\epsilon>0$, 
obtaining  a non-unitary representation of the $so(1,2)$ algebra which is  
equivalent to an unitary one. The case of $\epsilon=0$ is studied in the Sec. 6.

\section{Preliminaries}
\

Let us consider a static background with the natural frame $(t,x)$ in which the 
metric is $g_{\mu \nu}(x)$, with $\mu, \nu =0,1$. We shall assume that this is 
symmetric with respect to $x=0$ and we shall denote  $g=det(g_{\mu \nu})$. 
The domain, $D$, of the free motion observed by an observer situated at 
$x=0$ is that bounded by the observer's event horizon. In general, this is  
$D=(- x_{e}, x_{e})$ where $\pm x_{e}$ are either the finite points where the 
metric is singular or $\pm\infty$ in the case of the regular metrics.  
On this domain,  we shall define  the scalar field  $\phi$ of the mass 
$m$, supposing that this is minimally coupled with the gravitational field 
\cite{B1}. In the case of static backgrounds the  energy is conserved. 
Therefore,  the Klein-Gordon equation
\begin{equation}\label{(kg)}
\frac{1}{\sqrt{-g}}\partial_{\mu}\left(\sqrt{-g}g^{\mu\nu}\partial_{\nu}\phi
\right) + m^{2}\phi=0,
\end{equation}
admits a set of fundamental solutions (of positive and negative frequency) of 
the form
\begin{equation}\label{(sol)}
\phi_{E}^{(+)}=\frac{1}{\sqrt{2E}}e^{-iEt}U_{E}(x), \quad 
\phi^{(-)}=(\phi^{(+)})^{*}.
\end{equation}
These must be orthogonal with respect to the relativistic scalar product 
\cite{B1} 
\begin{equation}\label{(sp1)}
(\phi,\phi')=i\int_{D}dx\sqrt{-g}g^{00}
\phi^{*}\stackrel{\leftrightarrow}{\partial_{0}} \phi',
\end{equation}
which, in fact, reduces to the following scalar product of the wave functions 
$U$
\begin{equation}\label{(psc1)}
(U,U')=\int_{D}dx\mu(x)U^{*}(x)U'(x)
\end{equation}
where
\begin{equation}
\mu=\sqrt{-g}g^{00}.
\end{equation}

Now we observe that, according to (\ref{(sol)}),  the one-particle state space,  
$H$, coincides to that of the antiparticle. In the coordinate representation  
this is the space $L^{2}(D,\mu)$ of the  square integrable functions with 
respect to the scalar product (\ref{(psc1)}).  All these functions must 
satisfy the condition
\begin{equation}\label{(dom)}
\lim_{x\to \pm x_{e}}U(x)=0
\end{equation}
which is obvious for $x_{e}=\infty$. Moreover, when  $x_{e}$ is finite then 
the metric as well as the weight function $\mu$ are singular at $\pm x_{e}$ and, 
consequently, the condition (\ref{(dom)}) is also necessary.      
A set of wave functions, $U_{n}$, $n=0,1,..$, represents a  countable basis 
in $H$ if these are orthonormal, 
\begin{equation}\label{(10)}
(U_{n},U_{n'})=\delta_{n,n'}, 
\end{equation}
and satisfy the  completeness relation 
\begin{equation}
\sum_{n}{U_{n}}^{*}(x)U_{n}(x')=\frac{1}{\mu(x)}\delta(x-x').
\end{equation} 
The linear operators on $H$ will be denoted using boldface. They 
can be defined either by giving their matrix elements in  a countable 
basis  or as differential operators in the coordinate representation.
The most general  differential operator we shall use here will have the form
\begin{equation}\label{(opd)}
({\bf D}U)(x)=i\left[f(x)\frac{d}{dx}+h(x)\right]U(x) ,
\end{equation}
depending on two arbitrary real functions $f$ and $h$.
Its  adjoint with respect to the scalar product (\ref{(psc1)}) is
\begin{equation}\label{(opda)}
{\bf D}^{+}={\bf D}+i\left[\frac{1}{\mu}\frac{d(\mu f)}{dx}-2h\right]{\bf 1}
\end{equation}  
where  ${\bf 1}$ is the unit operator. 
Hereby, we see that for $h=\partial_{x}(\mu f)/2\mu$ the operator ${\bf D}$ is 
selfadjoint.

In general,  any (1+1)  static background admits a special natural 
frame, $(t,\hat x)$,  in which the metric  is  a  conformal transformation 
of the Minkowski flat metric. This  new frame can be obtained by changing the 
space coordinate   
\begin{equation}\label{(100)}
x\to \hat x = \int dx\mu(x) + const,
\end{equation}
so that 
\begin{equation}
\hat g_{00}(\hat x)=-\hat g_{11}(\hat x)=\sqrt{-\hat g(\hat x)} 
\end{equation}
and $\hat\mu(\hat x)=1$. Then, from (\ref{(opda)}) it results that 
the momentum operator $\sim i\partial_{\hat x}$ is selfadjoint. 
The state space $H$ is represented now by $L^{2}(\hat D)$ where 
$\hat D$ is the domain of the new space coordinate corresponding to $D$. It is 
obvious that both the spaces $L^{2}(D,\mu)$ and  $L^{2}(\hat D)$ come from 
the same coordinate representation of the space $H$ since the change of the 
continuous parameter of a generalized basis changes only the normalization scale.

\section{Relativistic oscillators}
\

Let us first  discuss the general case of $\epsilon >0$ and then turn to 
the limit  $\epsilon \to 0$. 
In the frames $(t,x)$, the metrics are given by (\ref{(m)}) where 
we have to change the parameters according to (\ref{(par2)}). 
For the models with  $\epsilon >0$ these metrics are singular at 
$\pm 1/\hat\omega$ so that $D=(-1/\hat\omega, 1/\hat\omega)$.  

The Klein-Gordon equation can be put in the form 
\begin{equation}\label{(kg1)}
\left(-(1-\hat\omega^{2}x^{2})\frac{d^2}{dx^2}+\hat\omega^{2}x\frac{d}{dx}  
+\frac{m^2}{\epsilon^{2}}\frac{\hat\omega^{2}x^{2}}
{1-\hat\omega^{2}x^{2}}\right)U(x)=  
({E}^{2}-m^{2})U(x)  
\end{equation}
while the weight function which defines the scalar product (\ref{(psc1)}) is 
\begin{equation}\label{(psc10)}
\mu(x)  =\frac{1}{\sqrt{1-\hat\omega^{2}x^{2}}}.
\end{equation}
Since the energy spectrum is countable, the energy eigenfunctions 
are the square integrable solutions  of (\ref{(kg1)}). These can be written in 
terms of hypergeometric functions as \cite{COTA}
\begin{equation}\label{(U1)}
U_{n}(x)=N_{n_{s},s}(1-\hat\omega^{2}x^{2})^{\frac{k}{2}}x^{s}F(-n_{s},k+s+n_{s},
s+\frac{1}{2}, \hat\omega^{2}x^{2}),
\end{equation}
where the parameter 
\begin{equation}\label{(p)}
k=\frac{1}{2}\left[1 + \sqrt{1+ 4\frac{m^{2}}{\epsilon^{2}\hat\omega^{2}}}
\right] > 1
\end{equation}\label{(wei)}
is the positive solution of the equation
\begin{equation}\label{(kk)}
k(k-1)=\frac{m^2}{\epsilon^{2}\hat\omega^2}.
\end{equation}
The quantum numbers, $n_{s}=0,1,2...$ and $s=0, 1$, 
can be embedded into the main quantum number $n=2n_{s}+s$. This will take 
even values if $s=0$ and odd values for $s=1$. Hence, the functions 
$U_{n}(x)$ are real polynomials of the degree $n$ in $x$, with the factor 
$(1-\hat\omega^{2} x^{2})^{k/2}$, which assures the condition (\ref{(dom)}). 
The normalization factors can be easily calculated  in
terms of Jacobi polynomials \cite{B2}. The result is
\begin{equation}\label{(normal)}
N_{n_{s}, s } = (-1)^{n_{s}}{( \hat\omega )^{s + {1\over{2}}} \over{
\sqrt{n_{s} !}}} (s + {1\over{2}})_{n_{s}} \left[ {(k  +  s + 2 n_{s})
\Gamma ( k +  s + n_{s} ) \over \Gamma (n_{s} +  s + { 1 \over 2 } )
\Gamma (n_{s} +  k + { 1 \over 2 } ) } \right]^{{ 1 \over 2 }} 
\end{equation}
where  we have used the notation $(z)_n = z (z + 1) ... (z + n - 1) $. Moreover, 
the normalized energy eigenfuncions can also be written in terms of associated 
Legendre polynomials \cite{PT2}.

The energy levels result from the quantization condition \cite{COTA}, 
\begin{equation}\label{(qcond)}
{E_{n}}^{2}-m^{2}\left(1- \frac{1}{\epsilon^2}\right)=\hat\omega^{2}(k+n)^{2},
\end{equation}
which gives
\begin{equation}\label{(el1)}
{E_{n}}^{2}=m^{2}+\hat\omega^{2}[2k(n+\frac{1}{2})+n^{2}], \quad 
n=0,1,2... .
\end{equation}  
for $\epsilon\not=1$, and
\begin{equation}\label{(el2)}
E_{n}=\hat\omega(k+n)
\end{equation}
in the case of the RHO \cite{M}, when $\epsilon=1$. 

We can conclude that our RO with $\epsilon >0$ are  systems of massive 
scalar particles confined to wells. Their properties are determined by three 
parameters, $m$, $\omega$ and $\epsilon$. This last one is our new parameter 
which gives the desired well width, $2/\epsilon\omega$, when the frequecy 
$\omega$ is fixed. It is interesting that all these parameters are concentrated 
in the expression of $k$ so that the eigenfunctions (\ref{(U1)}) depend only on 
$\hat\omega=\epsilon\omega$ and $k$ while the energies (\ref{(el1)}) involve 
all of them. Thus there is a  possibility to have RO with different energy 
spectra but having the same energy eigenfunctions.

\section{The relativistic P\" oschl-Teller problem}
\

Now we shall change the space coordinate according to (\ref{(100)}) where $\mu$ 
is given by (\ref{(psc10)}). We obtain 
\begin{equation}
\hat x=\frac{1}{\hat\omega}\arcsin\hat\omega x .
\end{equation}
In the new frame $(t, \hat x)$  the line element is
\begin{equation}
ds^{2}=\left(1+\frac{1}{\epsilon^2}\tan^{2}\hat\omega \hat x\right)(dt^{2}-
d\hat x^{2})
\end{equation} 
and $\hat D =(-\pi/2\hat\omega, \pi/2\hat\omega)$.  
The Klein-Gordon equation takes the form
\begin{equation}\label{(kg2)}
\left(-\frac{d^2}{d\hat x^{2}} +\frac{m^2}{\epsilon^2}\tan^{2}\hat\omega\hat x
\right)U_{n}(\hat x)=({E_{n}}^{2}-m^{2})U_{n}(\hat x).
\end{equation}
The second term of its left-hand side  can be rewritten using (\ref{(kk)}):  
\begin{equation}
V_{PT}(\hat x)=k(k-1)\hat\omega^2\tan^{2}\hat\omega \hat x.
\end{equation}
This will be called the relativistic (symmetric) PT 
potential since the solutions (\ref{(U1)}) in the new variable $\hat x$,  
\begin{equation}\label{(U2)}
U_{n}(\hat x)=N_{n_{s},s}\hat\omega^{-s} \cos^{k}\hat\omega\hat x 
\sin^{s} \hat\omega\hat x F(-n_{s}, k+s+n_{s},
s+\frac{1}{2},\sin^{2} \hat\omega \hat x),
\end{equation}
{\it coincide} with those given by the non-relativistic PT potential $V_{PT}/2m$.  
Of course, the energies, as well as the significance of the parameters, differ 
from those of the non-relativistic case.

On the other hand, we know that the functions (\ref{(U2)}) represent a complete 
set of orthonormal functions in $L^{2}(\hat D)$. This means that the set
$U_{n}, n=0,1,..$ is a countable basis in $H$, namely  the energy 
basis. Its shift operators \cite{PT2}      
\begin{eqnarray}
({\bf A}_{(+)}U_{n})(\hat x)&=&\frac{1}{\hat\omega\sqrt{2k}} 
[-\cos\hat\omega\hat x {d\over d\hat x} +
 \hat\omega \sin\omega\hat x ( k + n ) ]U_{n}(\hat x), \label{(270)}\\
({\bf A}_{(-)}U_{n})(\hat x)&=&\frac{1}{\hat\omega\sqrt{2k}} 
[\cos\hat\omega\hat x {d\over d\hat x} +
 \hat\omega \sin\omega\hat x ( k + n ) ]U_{n}(\hat x)\label{(280)}
\end{eqnarray}                                                
have the action
\begin{equation}\label{(290)}
{\bf A}_{(+)}U_{n}=C_{n}^{(+)}U_{n+1},\quad 
{\bf A}_{(-)}U_{n}=C_{n}^{(-)}U_{n-1},
\end{equation}
where 
\begin{eqnarray}
C_n^{(+)}&=& {1\over  \sqrt{2k}} \, \left( {(2 k + n)(k + n)
 \over  k + n + 1 }\right)^{1\over 2} \sqrt{n+1},\label{(250)}\\                                                 
C_n^{(-)}&=& {1\over  \sqrt{2k}} \, \left( {(2 k + n - 1)(k + n)
 \over  k + n - 1 }\right)^{1\over 2} \sqrt{n}.\label{(260)}
\end{eqnarray}
We note that according to (\ref{(opda)}) the shift operators are not related 
between them through  Hermitian conjugation, i.e. ${\bf A}_{(\pm)}^{+}\not=
{\bf A}_{(\mp)}$.
 
\section{Algebra}

\subsection{The differential operators}
\

Let us consider the position and momentum selfadjoint operators,  
\begin{equation}\label{(110)}
(\hat{\bf X}U)(\hat x)=\hat x U(\hat x), \quad
(\hat{\bf P}U)(\hat x)=i\frac{dU(\hat x)}{d\hat x},
\end{equation}
in the frame $(t, \hat x)$, with the commutation rule 
\begin{equation}\label{(120)}
[\hat {\bf P}, \hat {\bf X}]=i{\bf 1}.
\end{equation}   
Then, from (\ref{(kg2)})  we see that the energy square operator is  
\begin{equation}\label{(epx)}
{\bf E}^{2}=m^{2}{\bf 1}+\hat{\bf P}^{2}+\frac{m^2}{\epsilon^2}
\tan^{2}\hat\omega \hat{\bf X}.
\end{equation}
Its form suggests us to introduce the  pair of adjoint operators \cite{PT3}
\begin{eqnarray}
{\bf \tilde a}=\frac{1}{\hat\omega\sqrt{2k}}\left(-i\hat{\bf P}+
k\hat\omega \tan \hat\omega \hat{\bf X}\right),\label{(300)}\\
{\bf \tilde a}^{+}=\frac{1}{\hat\omega\sqrt{2k}}\left(i\hat{\bf P}+
k\hat\omega \tan \hat\omega \hat{\bf X}\right).\label{(310)}
\end{eqnarray}
which satisfy the commutation relation
\begin{equation}
[{\bf \tilde a}, {\bf \tilde a}^{+}]= {\bf 1}+\frac{1}{2k}({\bf \tilde a}^{+}+
{\bf \tilde a})^{2}
\end{equation}
and allow to write
\begin{equation}
{\bf E}^{2}=m^{2}{\bf 1}+ 2k\hat\omega^{2}\left({\bf \tilde a}^{+}{\bf \tilde a} 
+\frac{1}{2} {\bf 1} \right) .
\end{equation}
The commutation relations of ${\bf E}^2$ with $\hat{\bf X}$ and $\hat{\bf P}$ 
can also be calculated from (\ref{(120)}) and (\ref{(epx)}). Another 
exercise is to replace the position operator by the effective position operator 
\begin{equation}
\hat{\bf X}_{ef}= \frac{1}{\hat\omega}\tan\hat\omega\hat{\bf X}
\end{equation} 
in order to recover the familiar formulas
\begin{equation}
\hat{\bf X}_{ef}=\frac{1}{\hat\omega\sqrt{2k}}({\bf \tilde a}^{+}+
{\bf \tilde a}), \qquad 
\hat{\bf P}=-i\hat\omega\sqrt{\frac{k}{2}}({\bf \tilde a}^{+}-
{\bf \tilde a}). 
\end{equation}

\subsection{The non-differential operators}
\

However, there are other operators which are not differential. The analysis 
of their structure  can be done by  using the operators of the standard 
oscillator algebra, ${\bf a}$, ${\bf a}^{+}$ (with $[{\bf a}, 
{\bf a}^{+}]={\bf 1}$) and ${\bf N}={\bf a}^{+}{\bf a}$,  which 
can be defined  in the energy basis as follows: 
\begin{equation}\label{(sta)} 
{\bf a}^{+}U_{n}=\sqrt{n+1}U_{n+1}, \quad   
{\bf a}U_{n}=\sqrt{n}U_{n-1}, \quad
{\bf N}U_{n}=nU_{n}.
\end{equation}
Now the quantization condition (\ref{(qcond)}) becomes
\begin{equation}
{\bf E}^{2} =
\hat\omega^{2}[({\bf N}+k{\bf 1})^{2}+(\epsilon^{2}-1)k(k-1){\bf 1}],
\end{equation}
while  the shift operators  can be put in the form 
\begin{eqnarray}
{\bf A}_{(+)}&=& 
(\cos\hat\omega\hat{\bf X}) {\bf \tilde a}^{+}+
\frac{1}{\sqrt{2k}}(\sin\hat\omega\hat{\bf X}){\bf N},\\
{\bf A}_{(-)}&=& 
(\cos\hat\omega\hat{\bf X}) {\bf \tilde a}+
\frac{1}{\sqrt{2k}}(\sin\hat\omega\hat{\bf X}){\bf N}.
\end{eqnarray}
Furthermore, from Eqs. (\ref{(270)}) - (\ref{(260)}) it results that these 
operators can be expressed  in terms of the generators of the standard 
oscillator algebra as
\begin{equation}\label{(uj)}
{\bf A}_{(+)}={\bf a}^{+}w_{(+)}({\bf N}), \quad 
{\bf A}_{(-)}=w_{(-)}({\bf N}) {\bf a},
\end{equation}
where
\begin{equation}
w_{(+)}({\bf N})=w_{(-)}({\bf N})\frac{ {\bf N}+k{\bf 1}}{ {\bf N}+
k{\bf 1}+{\bf 1}}=\left[ \frac{({\bf N}+2k{\bf 1})({\bf N}+k{\bf 1})}{2k({\bf N}+
k{\bf 1}+{\bf 1})} \right] ^{\frac{1}{2}}.
\end{equation}  
Hereby, we obtain the commutation relations 
\begin{equation}
[{\bf A}_{(-)}, {\bf A}_{(+)}]= {\bf 1}+\frac{1}{k} {\bf N}, \quad
[{\bf N}, {\bf A}_{(\pm)}]=\pm {\bf A}_{(\pm)}
\end{equation}
and the identity
\begin{equation}\label{(kgo)}
2k{\bf A}_{(+)}{\bf A}_{(-)}={\bf N}[{\bf N}+(2k-1){\bf 1}]
\end{equation}
which is nothing else than the operator form of the Klein-Gordon equation, as 
it results fom (\ref{(270)}), (\ref{(280)}) and (\ref{(sta)}).

Let us observe that the operators $\sqrt{2k}{\bf A}_{(+)}$, 
$\sqrt{2k}{\bf A}_{(-)}$ and ${\bf N}+k{\bf 1}$ are the generators of a 
non-unitary representation of the  $so(1,2)$ algebra. This is (non-unitary) 
equivalent with the unitary representation of the lowest weight $k$, which 
has the generators
\begin{eqnarray}
{\bf K}_{(+)}&=&{\bf K}_{1}+i{\bf K}_{2}=\nonumber\\
&=&\sqrt{2k}({\bf N}+k{\bf 1})^{\frac{1}{2}}
{\bf A}_{(+)} ({\bf N}+k{\bf 1})^{-\frac{1}{2}} = 
{\bf a}^{+}({\bf N}+2k{\bf 1})^{\frac{1}{2}}\\ 
{\bf K}_{(-)}&=&{\bf K}_{1}-i{\bf K}_{2}=\nonumber\\
&=&\sqrt{2k}({\bf N}+k{\bf 1})^{\frac{1}{2}}
{\bf A}_{(-)} ({\bf N}+k{\bf 1})^{-\frac{1}{2}} = 
({\bf N}+2k{\bf 1})^{\frac{1}{2}}{\bf a}\\ 
{\bf K}_{3}&=& {\bf N} +k{\bf 1}.
\end{eqnarray}
The Casimir operator
\begin{equation}
{{\bf K}_{3}}^{2}-{{\bf K}_{1}}^{2} - {{\bf K}_{2}}^{2}= k(k-1){\bf 1}
\end{equation}
is an alternative form of the Klein-Gordon operator (\ref{(kgo)}). 

Finally we must specify that the  operators $\hat{\bf X}$ and $\hat{\bf P}$ 
can be expressed  in terms of ${\bf a}$ and ${\bf a}^{+}$ by using the Eqs. 
(\ref{(270)}), (\ref{(280)}) and (\ref{(uj)}). However, we can introduce 
other coordinate and momentum operators corresponding to all the  natural 
frames we desire to choose. Obviously, all these operators are analytic functions 
of $\hat{\bf X}({\bf a}, {\bf a}^{+})$ and $\hat{\bf P}({\bf a}, {\bf a}^{+})$. 
Thus it results that the whole operator algebra of the RO is freely generated 
by ${\bf a}$ and ${\bf a}^{+}$ only.

\section{The limit $\epsilon \to 0$}
\

The case of $\epsilon=0$ can be solved separately \cite{COTA}. Here we have 
$x=\hat x$ and $x_{e}\to \infty$ so that, in the coordinate representation,  
$H$ will appear as  $L^{2}(R)$. The solutions of the  
Klein-Gordon equation,
\begin{equation}\label{(kg4)}                                                 
- \frac{d^{2} U_{n}^{0}}{dx^2} + m^2 \omega^2 x^2 U_{n}^{0} = 
({E_{n}}^2 - m^2 ) U_{n}^{0},
\end{equation}
 coincide with the familiar energy eigenfunctions of the NRHO, while 
the energy spectrum is given by
\begin{equation}\label{(s1)}
{E_{n}}^{2}=m^{2}+2m\omega(n+\frac{1}{2}), \quad n=0,1,2,...
\end{equation} 

On the other hand, we have shown \cite{COTA} that the solutions of our RO are 
continuous in $\epsilon=0$ in the sense that the limit of the energy 
eigenfunction (\ref{(U1)}) for $\epsilon\to 0$, calculated up to the 
normalization factors, gives just the solutions of (\ref{(kg4)}).  
Now we can convince ourselves, that the normalization factors (\ref{(normal)}) 
also behave correct in the limit $\epsilon \to 0$, giving the usual normalization 
factors of the NRHO energy eigenfunctions, $U^{0}_{n}$, which should satisfy 
$(U^{0}_{n},U^{0}_{n'})=\delta_{nn'}$. Indeed, by taking into account that in 
this limit we have $\hat\omega\to 0$, $k\to\infty$ but $\epsilon^{2}k\to 
m/\omega$, and by using the asymptotic form of the functions $\Gamma(z)$ for 
large $z$ \cite{B2}, we find that
\begin{eqnarray}
\lim_{\epsilon \to 0} U_{n}(x) = 
\left(\frac{m\omega}{\pi}\right)^{\frac{1}{4}} 
\frac{(-1)^{n_{s}}(s+1)}{2^{n_{s}+\frac{s}{2}}n_{s}!}\sqrt{(2n_{s}+s)!} \times \nonumber\\
e^{-m\omega x^{2}/2}(\sqrt{m\omega} x)^{s}F(-n_{s},s+\frac{1}{2}, 
m\omega x^{2})=  \label{(1000)}\\
=\left(\frac{m\omega}{\pi}\right)^{\frac{1}{4}} 
\frac{1}{\sqrt{n! 2^{n}}}e^{-m\omega x^{2}/2}H_{n}(\sqrt{m\omega}x)
=U_{n}^{0}(x) ,\nonumber
\end{eqnarray}
where $H_{n}$ are the Hermite polynomials and $n=2n_{s}+s$ as defined above. 
We note that this result justifies the choice of the phase factor 
$(-1)^{n_{s}}$ of (\ref{(normal)}).

The behavior of the energy eigenfunctions suggests us  to derive the 
operator algebra in the case of $\epsilon=0$ as the limit of the algebra 
obtained in the previous section. We observe that $w_{(\pm)}({\bf N})\to  
{\bf 1}$ when $\epsilon \to 0$ so that
\begin{equation}
\lim_{\epsilon \to 0}{\bf A}_{(+)}=
\lim_{\epsilon \to 0}{\bf \tilde a}^{+}={\bf a}^{+},\quad   
\lim_{\epsilon \to 0}{\bf A}_{(-)}=
\lim_{\epsilon \to 0}{\bf \tilde a}={\bf a}  . 
\end{equation}   
Furthermore, from (\ref{(270)}) and (\ref{(280)}) we see that the operators 
${\bf a}$ and ${\bf a}^{+}$ become  differential operators of the form
\begin{eqnarray}                                                                                                  
({\bf a}U)(x) = {1 \over \sqrt{2 m \omega}} ({d \over d x} + 
m \omega x ) U(x),\\
({\bf a}^{+}U)(x) = {1\over \sqrt{2 m \omega}} ( - {d \over d x} + 
m \omega x)U(x).
\end{eqnarray}
This means that the position and the momentum operators can be written  as
in the non-relativistic case, 
\begin{equation}
{\bf X}= \frac{1}{\sqrt{2m\omega}}({\bf a}^{+}+{\bf a}), \quad
{\bf P}=-i\sqrt{\frac{m\omega}{2}} ({\bf a}^{+}-{\bf a}).
\end{equation}
Hence, the conclusion is that the $so(1,2)$ algebra of the models with 
$\epsilon>0$ degenerates in the usual NRHO algebra when 
$\epsilon\to 0$. Simultaneously, the Klein-Gordon operator (\ref{(kgo)}) 
becomes ${\bf a}^{+} {\bf a}={\bf N}$. Thus, for 
$\epsilon=0$ the energy eigenfunctions as well as the shift operators 
are the same as those of the NRHO. The unique difference is the formula 
of the energy levels which gives the relativistic energy square operator 
\begin{equation}\label{(500)}
{\bf E}^{2}=m^{2}{\bf 1}+2m\omega({\bf N}+\frac{1}{2}{\bf 1}).
\end{equation}
and new commutation rules for ${\bf E}^{2}$ with ${\bf X}$ or ${\bf P}$. 

\section{Comments}
\

In this article we have studied the properties of the fundamental solutions 
of the Klein-Gordon equations of the RO by using the traditional methods of the 
coordinate representation. This allowed us to study the form of the energy 
eigenfunctions in two natural frames. The first one, $(t,x)$, is important since 
here the classical equations of motion of the RO look like that of the NRHO. 
This indicates that the physical meaning of the relativistic behavior of the RO 
could be better point out in this frame. The other frame, $(t, \hat x)$, offers 
the advantage of the simplest form of the Klein-Gordon equation (\ref{(kg2)}). 
We have shown that for $\epsilon >0$ this is the relativistic version of the  
PT system. Moreover, we have seen that in the limit $\epsilon\to 0$ both these 
frames coincide while the solutions of the Klein-Gordon equation become just 
those of the NRHO. Thus we can conclude that, at least in the frame $(t, \hat x)$,  
the space behavior  of the RO remains very closed to that of several 
non-relativistic systems. However, the operators of physical interest have new 
properties and specific and consistent, although quit complicated, commutation 
relations. We must specify that our results concerning the commutation 
rules of the energy, position and momentum operators differ from those predicted 
in the algebraic  approach of the RHO \cite{A1}. What is remarkable here is that 
we have recovered on an other way the $so(1,2)$ algebra, for all the RO with 
$\epsilon>0$, so that the Klein-Gordon operator should be just its Casimir 
operator. As mentioned, all these systems are of massive scalar particles 
confined to wells having the width $2/\epsilon\omega$ in the frame $(t, x)$ 
or $\pi/\epsilon\omega$ in the frame $(t, \hat x)$.

Finally we note that all the results we have obtained could be the starting 
point of the construction of the  quantum field theory of the scalar field 
$\phi$. This must be defined on $H$ with values in the field operator algebra 
by introducing suitable  creation and annihilation operators. Moreover, all the 
operators we have discussed here will generate  the one-particle operators of 
the quantum field theory. Then the physical meaning of these operators will be 
better understood because of the possibility of analyzing their time evolution 
with the help of the Hamiltonian operator. In our opinion, in this way one could 
obtain new answers in some  sensitive problems such as that related to the 
definition of the relativistic position and momentum operators.

\end{document}